\documentclass[
reprint,
superscriptaddress,
nofootinbib,
amsmath,amssymb,
aps,
prl,
floatfix,
twocolumn,
notitlepage
]{revtex4-2}

\usepackage[utf8]{inputenc}
\usepackage[T1]{fontenc}
\usepackage{amsmath}
\usepackage{amssymb}
\usepackage{graphicx}
\usepackage{subcaption}
\usepackage{booktabs}
\usepackage[table]{xcolor}
\usepackage[normalem]{ulem}
\usepackage{microtype}

\begin{document}


\title{\large\bfseries 
How Rich is the Rich-Club? \\Super Rich-Club Organization in Complex Networks
}

\author{Lucía S. Ramirez}
\affiliation{Departament de Física de la Matèria Condensada, Universitat de Barcelona, Martí i Franquès 1, 08028, Barcelona, Spain}
\affiliation{Universitat de Barcelona Institute of Complex Systems (UBICS), Barcelona, Spain}

\author{Laia Barjuan}
\affiliation{Departament de Física de la Matèria Condensada, Universitat de Barcelona, Martí i Franquès 1, 08028, Barcelona, Spain}
\affiliation{Universitat de Barcelona Institute of Complex Systems (UBICS), Barcelona, Spain}

\author{M. \'Angeles Serrano}
\affiliation{Departament de Física de la Matèria Condensada, Universitat de Barcelona, Martí i Franquès 1, 08028, Barcelona, Spain}
\affiliation{Universitat de Barcelona Institute of Complex Systems (UBICS), Barcelona, Spain}
\affiliation{Institució Catalana de Recerca i Estudis Avançats (ICREA), Passeig Lluís Companys 23, E-08010 Barcelona, Spain}

\date{\today}

\begin{abstract}
\noindent
The rich-club coefficient is used in complex networks to quantify the tendency of progressively higher-degree nodes to be more densely connected than expected by chance. However, it does not capture whether high degree nodes towards the core are actually more interconnected than the original network.
Here, we jointly analyze the rich-club coefficient and an enrichment coefficient, which quantifies whether the internal average degree of progressively higher-degree subgraphs is greater than that of the original network, despite the decreasing number of nodes. We apply this framework to real and synthetic networks. Our results reveal three distinct regimes of core organization: rich-club organization without enrichment, enrichment without rich-club organization, and a combination of both, which we call Super Rich-Club organization. In  the latter regime, high-degree nodes form cohesive and unusually dense cores that support efficient communication. Our results show that characterizing core organization across structural scales requires considering both relative density and internal connectivity, revealing a stronger form of core organization in real systems with implications for network structure, spectral properties that govern dynamical processes, and communicability.
\end{abstract}

\maketitle


Networks provide an abstract yet powerful representation of complex systems, in which nodes denote the system's components and edges encode their interactions. In many real-world networks, a small fraction of nodes---the hubs---exhibit a large number of connections and play a key role in shaping global structure and dynamics. These nodes can strongly influence processes such as robustness, information flow, and spreading dynamics~\cite{albert2002statistical}. Their influence, however, depends not only on their individual degree but also on how they are interconnected. Degree heterogeneity alone is therefore insufficient to characterize the organization of network cores. Indeed, networks with similar degree distributions may differ substantially in how their most connected nodes are organized, with important implications for their structural organization and dynamics~\cite{song2006origins,castellano2012competing,castellano2017relating}. 

In particular, the rich-club (RC) effect~\cite{zhou2004rich,colizza2006detecting} characterizes whether high-degree nodes form denser cores than expected in degree-reserving randomized counterparts. RC organization has been identified across social, technological, and biological networks, and the concept has been extended to weighted~\cite{serrano2008rich,opsahl2008prominence,zlatic2009rich}, hierarchical~\cite{mcauley2007richclub}, and temporal networks~\cite{pedreschi2022temporal}. The existence of a rich club can have important consequences for system-level properties such as routing efficiency, robustness, and the emergence of oligarchic structures. For example, in social networks it has been used to characterize elite structures in scientific funding and collaboration~\cite{ma2015anatomy,nakajima2023higher}, and in brain and neuronal networks it has been linked to high-capacity communication and integrative processing~\cite{vandenheuvel2011richclub,nigam2016rich,zheng2024rich,riedel2022}. 

However, the RC does not capture how the actual level of interconnectivity among nodes evolves as progressively stricter degree thresholds are applied to construct the cores. While it quantifies whether high-degree nodes are more densely connected than expected from a degree-preserving null model, it does not determine  the actual level of internal connectivity of the resulting hub core relative to that of the original network. This distinction is especially important in sparse real-world networks. In such networks, the density of degree-thresholded subgraphs may increase simply because nodes are removed faster than edges, even if their internal average degree stays constant or decreases. Consequently, a positive RC signal does not necessarily imply that the high-degree core is becoming more internally connected; it only indicates that it is denser than expected relative to a randomized counterpart. Two networks may therefore exhibit similar RC profiles while differing substantially in the evolution of their internal average degree along the degree filtration and hence in core organization. This limitation motivates the need for a complementary measure that captures the cohesiveness of subgraphs in the nested hierarchy.

Here, we analyze the flow of the internal average degree of the nested sequence of subgraphs in the degree-threshold filtration in combination with the RC effect in real and synthetic networks. We say that a subgraph displays \emph{enrichment} when its internal average degree exceeds that of the original network. We show that this behavior is distinct from, and complementary to, the RC effect: networks may display RC organization with or without enrichment, and enrichment may also occur in the absence of a RC signal. We refer to networks that display both enrichment and RC organization as exhibiting \emph{Super Rich-Club} (SRC) organization. In such networks, high-degree nodes are not only more densely connected than expected by chance, but also form more cohesive cores with higher internal average degree than the original network. We show that this distinction is associated with stronger concentration of connectivity within the high-degree core and with a larger response to core perturbations in terms of spectral properties of the adjacency matrix and in network communicability. These results indicate that enrichment captures a form of core organization with structural and dynamical consequences beyond RC ordering alone.

\section*{Rich Club and Enrichment:\\ Connectivity in Network Cores}

Here, we briefly review RC organization and introduce enrichment as a complementary measure of connectivity within degree-thresholded network subgraphs. While the RC coefficient assesses whether high-degree nodes are more densely connected than expected from the degree sequence, enrichment captures how the actual internal connectivity of the filtered subgraphs evolves along the degree-threshold filtration.


\begin{figure}
    \centering
    \includegraphics[width=\linewidth]{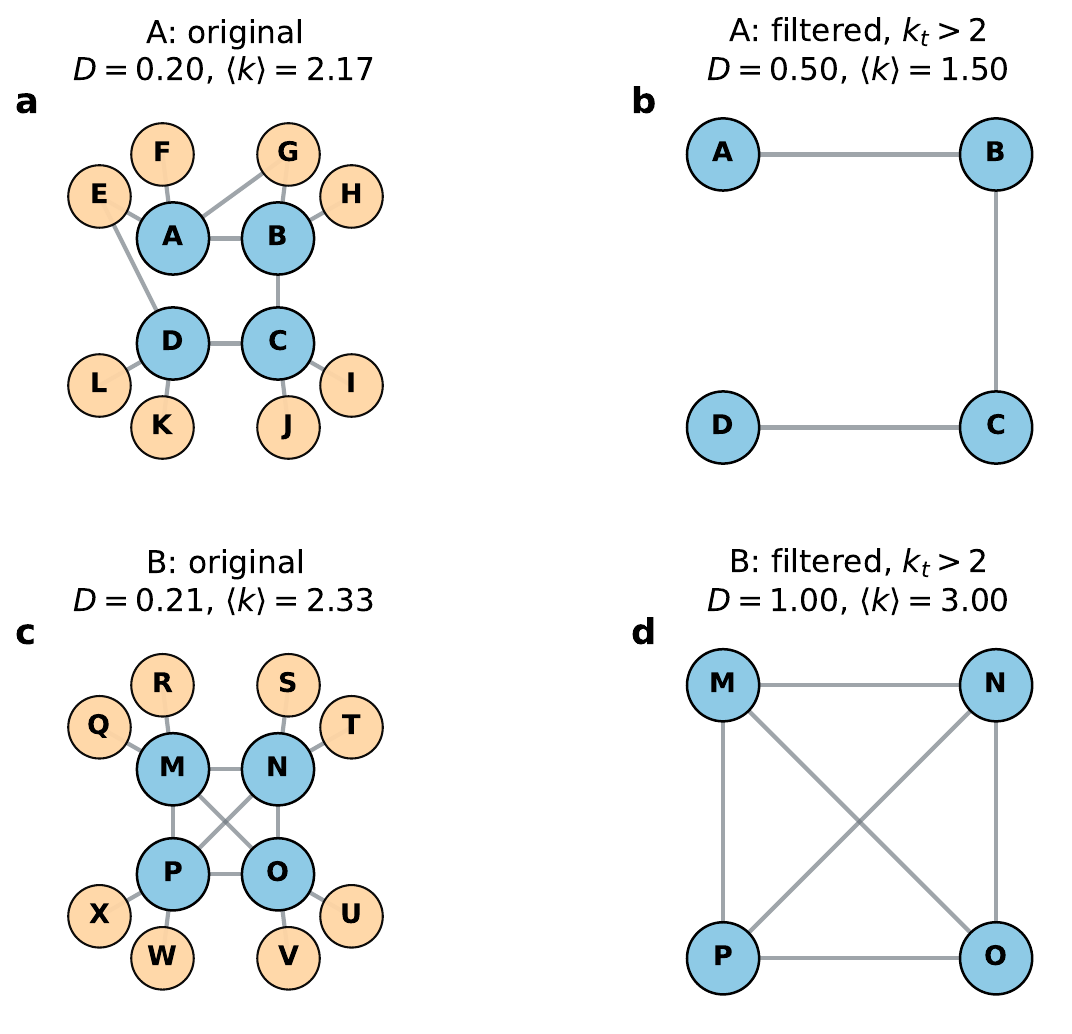}
   \caption{
    \textbf{Effect of degree-threshold filtration on the average degree. }
Sketch of two networks under degree-threshold filtration for $k_t = 2$. In both cases, the density of the resulting subgraphs increases; however, this does not necessarily imply an increase in their average degree. In (a), filtering increases the density but reduces the average degree relative to the original network, indicating the absence of enrichment. In (b), filtering increases both the density and the average degree, revealing the presence of enrichment.} \label{fig:sketch}
\end{figure}

\paragraph{RC organization.} 

The non-normalized RC coefficient quantifies the density of connections among nodes whose degree exceeds a given threshold~\cite{zhou2004rich}, providing a way to characterize whether progressively higher-degree nodes form increasingly dense subgraphs along a degree-threshold filtration. Specifically, a degree filtration generates a nested hierarchy of subgraphs $\mathcal{G}_{k_t}$ formed by nodes with $k>k_t$. For each threshold $k_t$, the non-normalized RC coefficient is
\begin{equation}
\phi_{\mathrm{nn}}(k_t)
=
\frac{2E_{>k_t}}
{N_{>k_t}(N_{>k_t}-1)},
\end{equation}
where $N_{>k_t}$ and $E_{>k_t}$ are, respectively, the number of nodes and edges in $\mathcal{G}_{k_t}$.

Because high-degree nodes are more likely to be densely connected simply as a consequence of their degree, $\phi_{\mathrm{nn}}(k_t)$ must be compared with a degree-preserving null model~\cite{colizza2006detecting}. The normalized RC coefficient is
\begin{equation}
\phi(k_t)
=
\frac{\phi_{\mathrm{nn}}(k_t)}
{\phi_{\mathrm{rand}}(k_t)},
\end{equation}
where $\phi_{\mathrm{rand}}(k_t)$ is the expected value in a degree-preserving randomized networks. Values $\phi(k_t)>1$
indicate that nodes in the corresponding subgraph are more densely connected than
expected from the degree sequence alone
~\cite{colizza2006detecting,vandenheuvel2011richclub,vandenheuvel2013network}.
Hereafter, we refer to this normalized coefficient as the RC coefficient.

\begin{figure*}[th!]
    \centering
    \includegraphics[width=\linewidth]{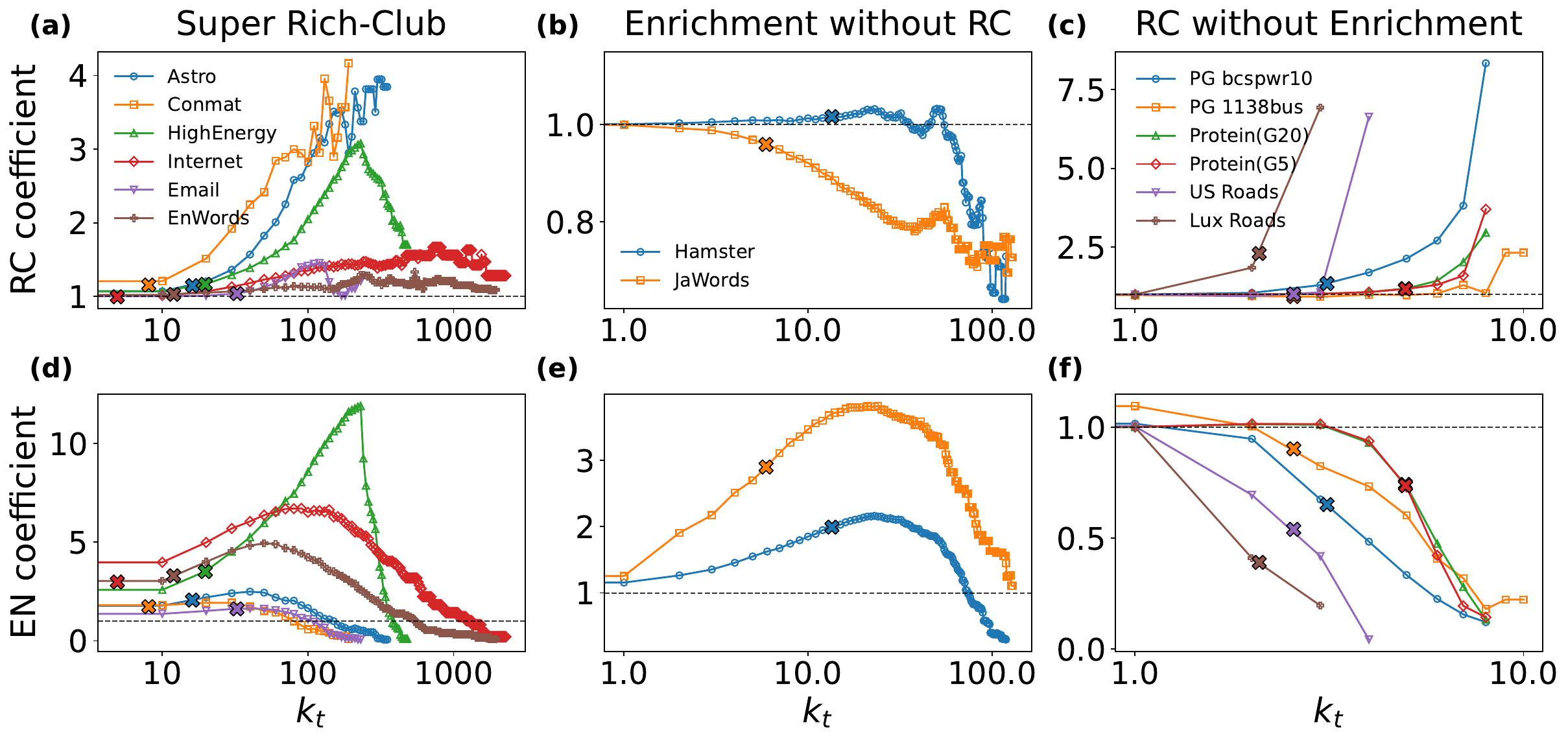}
    \caption{
    \textbf{ RC organization and enrichment in real networks.}
    The three columns correspond to the three structural regimes:
    \emph{Super Rich Club} (left),
    \emph{Enrichment without Rich Club} (middle), and
    \emph{Rich Club without Enrichment} (right).
    The top row shows the normalized rich-club coefficient,
    $\phi$, as a function of the degree threshold $k_t$,
    while the bottom row shows the enrichment coefficient, $\epsilon$, for the same networks.
    The crosses indicate the original mean degree $\langle k\rangle$ of each network.
    }
    \label{fig:RC_realnet}
\end{figure*}

\paragraph{Enrichment.}
In sparse networks---the most common situation in real-world systems---the average degree $\left<k\right>$ remains bounded while link density scales with system size as $N^{-1}$ and tends to zero as the network grows.
Along the flow of the degree-threshold filtration, the density of subgraphs $\mathcal{G}_{k_t}$ may increase simply because the number of nodes decreases faster than the number of edges, while the internal average degree $\langle k \rangle_{{k_t}}$ remains constant or even decreases. The RC coefficient may thus indicate a rich club effect, since it only reflects whether subgraphs are denser than their randomized counterparts, while the internal average degree of subgraphs progressively decreases.

To overcome this limitation, we propose to complement the RC coefficient with the study of the evolution of the internal average degree of the nested subgraphs, 
\begin{equation}
\langle k \rangle_{>k_t} = \frac{2 E_{>k_t}}{N_{>k_t}}.
\end{equation}
We quantify the increase in internal connectivity relative to the average degree of the original network through the \textit{enrichment coefficient},
\begin{equation}
\epsilon(k_t) =
\frac{\langle k \rangle_{>k_t}}{\langle k \rangle}.
\end{equation}
When the internal average degree of the degree-filtered subgraph exceeds that of the original network, $\epsilon(k_t) > 1$, we say that the network is in an \textit{enrichment regime} at threshold $k_t$. In this regime, the nodes surviving the filtering process are, on average, more interconnected among themselves than those in the original network, as illustrated by the toy examples in Fig.~\ref{fig:sketch}. The structural condition under which filtering produces enrichment is derived in Methods, under "Structural condition for enrichment".

Note that the enrichment coefficient normalized using degree-preserving randomization, $\epsilon_{\text{norm}}(k_t)$, is equivalent to the normalized rich club coefficient $\phi(k_t)$ and 
both expressions reduce to the same function 
\begin{equation}
\epsilon_{\text{norm}}(k_t) = \phi(k_t) = \frac{E_{>k_t}}{E^{\mathrm{rand}}_{>k_t}}.
\end{equation}

\section*{Super Rich-Club Organization of Network Cores} \label{sec:regimes}
The joint analysis of the RC and enrichment reveals that networks can be systematically classified according to how density and average degree evolve in the degree-threshold filtration, defining distinct structural scenarios that capture different organizations of the network core. Networks can display only enrichment, only RC organization or both. We define networks that display both enrichment and RC organization as exhibiting \emph{Super Rich-Club} (SRC) organization. In such networks, subgraphs of high-degree nodes are not only more densely connected than expected by chance, but also have a higher internal average degree than the original network. This implies the presence of a core that is simultaneously denser than expected under the null model and structurally cohesive. 

The three structural configurations are observed across real networks, as illustrated in Fig.~\ref{fig:RC_realnet}. We study publicly available data sets whose main characteristics can be found in Methods. A larger explanation is given in the Supplementary Information.

\paragraph{Super Rich-Club.}
The left column of Fig.~\ref{fig:RC_realnet} corresponds to real networks that display both a RC effect ($\phi>1$) and enrichment ($\epsilon>1$). High-degree nodes are not only more densely connected than expected by chance, but also form subgraphs whose internal average degree exceeds that of the original network, resulting in a highly interconnected core. Examples include citation networks, the Internet, and the English word-adjacency network.
\begin{figure*}
    \centering
    \includegraphics[width=\linewidth]{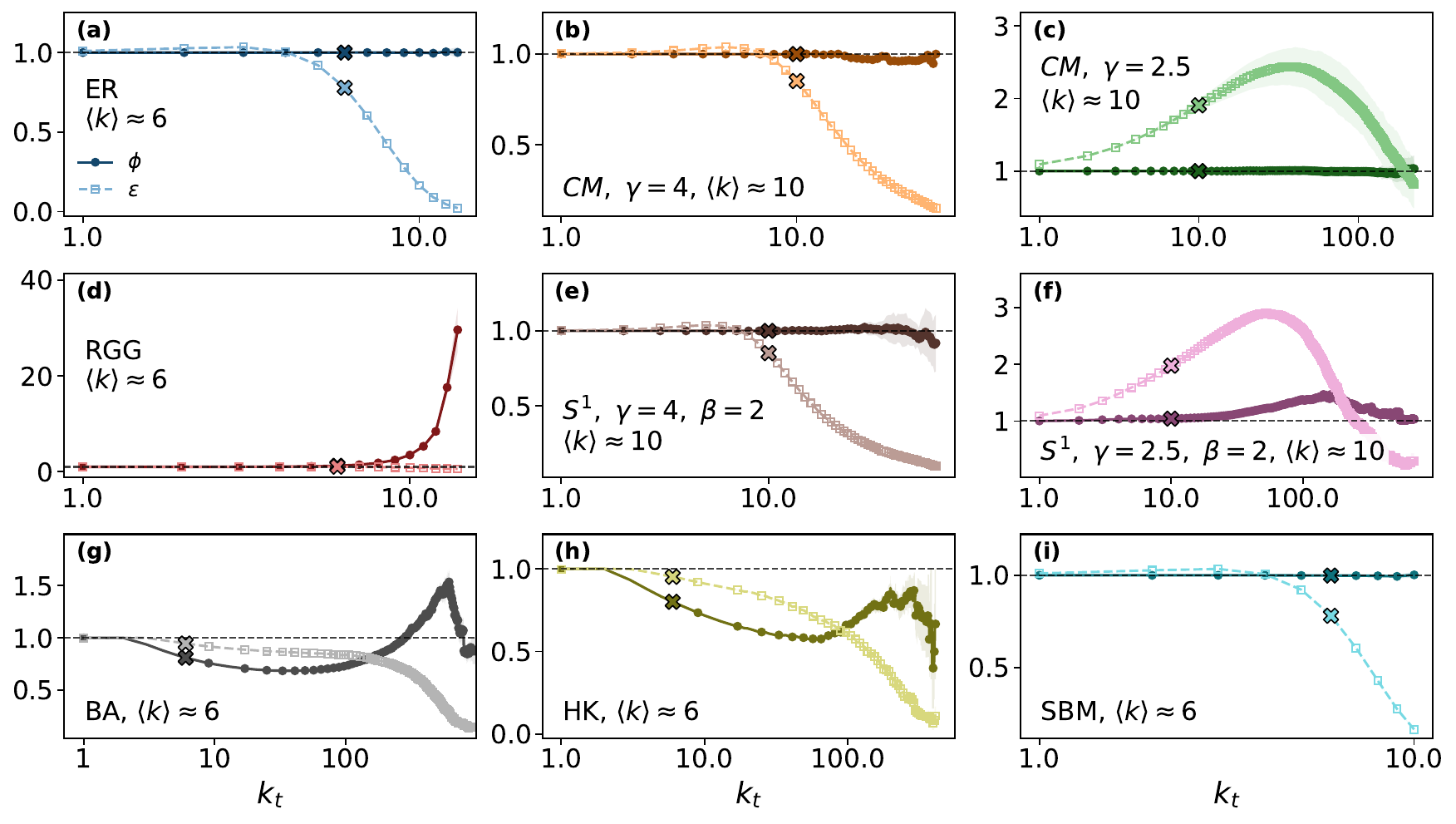}
        \caption{\textbf{Rich-club organization and enrichment in synthetic networks.}
    The normalized rich-club coefficient $\phi$ and the enrichment coefficient $\epsilon$ are shown as functions of the degree threshold $k_t$ for different synthetic network models with $N = 10^4$ nodes.
    Solid circles correspond to $\phi$, while open squares correspond to $\epsilon$.
    Shaded bands indicate sample-to-sample fluctuations across $100$ independent realizations.
    Crosses mark the mean degree $\langle k\rangle$ of each network. 
    }   \label{fig:RC_synthetic}
\end{figure*}

\paragraph{Enrichment without RC.}
Real networks in the middle column of Fig.~\ref{fig:RC_realnet} display enrichment ($\epsilon>1$) without a significant RC effect, or even with an anti-RC tendency ($\phi \leq 1$). 
Filtering isolates subgraphs whose internal average degree exceeds that of the original network, and therefore identifies an enriched core. However, despite this increased internal connectivity, the density of the core may still be consistent with, or even lower than, that expected from the degree sequence. In the latter case, the network exhibits enrichment despite a relative tendency toward hub–hub repulsion.

\paragraph{RC without enrichment.}
Real networks in the right column of Fig.~\ref{fig:RC_realnet} exhibit a RC effect ($\phi > 1$) but no enrichment ($\epsilon \leq 1$). High-degree nodes form subgraphs more densely connected than expected by chance;
however, the internal average degree tends to decrease as the core becomes more selective. This behavior is typical of infrastructure and protein interaction networks, which tend to have low mean degree and relatively homogeneous degree distributions.

To understand the mechanisms sustaining the different scenarios observed in real data, we analyzed a set of generative models widely used in the study of social, biological, and technological systems. Figure~\ref{fig:RC_synthetic} shows $\phi$ and $\epsilon$ for synthetic networks generated with the Erdős--Rényi (ER) model~\cite{erdos1959random}, the configuration model (CM)~\cite{molloy1995critical}, the $\mathbb{S}^1$ model~\cite{serrano2008selfsimilarity}, the Barabási--Albert (BA) model~\cite{barabasi1999emergence}, the Holme--Kim (HK) model~\cite{holme2002growing}, a stochastic block model (SBM)~\cite{holland1983sbm}, and random geometric graphs (RGG)~\cite{penrose2003random}. Further details on network construction and model parameters provided in Methods and Supplementary Information.

As reported before, uncorrelated networks such as ER or CM do not display a RC effect~\cite{colizza2006detecting}, but their enrichment behaviors differ. In ER networks, the internal average degree of the filtered subgraphs decreases monotonically with the degree threshold $k_t$. In contrast, CM networks show a dependence on the exponent $\gamma$ of the degree distribution: for $\gamma=3.5$, $\epsilon(k_t)$ decreases along the filtration, whereas for $\gamma=2.5$ it remains above one over almost the entire range. This can be understood from the scaling of degree-thresholded subgraphs in scale-free networks, $\langle k\rangle(k_t)\sim k_t^{3-\gamma}\langle k\rangle$~\cite{serrano2008selfsimilarity}, which implies that the internal average degree increases respect to the original one with $k_t$ when $\gamma<3$ and decreases when $\gamma>3$. 

Spatial constraints lead to a different scenario. Random geometric graphs (RGG) display an increasing RC coefficient at large $k_t$, while $\epsilon(k_t)$ remains below one. This is consistent with observations in spatially embedded infrastructure networks such as power grids and road networks in Fig.~\ref{fig:RC_realnet}. 
The $\mathbb{S}^1$ model~\cite{serrano2008selfsimilarity} provides a contrasting example. Its enrichment behavior follows the same scaling argument as the CM: for $\gamma=4$ the internal average degree decreases along the filtration, whereas for $\gamma=2.5$ it increases~\cite{serrano2008selfsimilarity}. In addition, the $\mathbb{S}^1$ model with heterogeneous degrees displays a RC effect, with the case $\gamma=2.5$ showing a clear SRC structure. 

The remaining models do not reproduce the RC or enrichment behavior in the parameter ranges analyzed here. Although BA, HK, and SBM networks incorporate ingredients commonly associated with real systems, such as preferential attachment, clustering, or community structure, these mechanisms are not sufficient on their own to generate both a RC effect and enrichment.

Both RC organization and enrichment are observed across real networks, yet they are not easily reproduced by standard synthetic models. The RC effect itself is known to be highly non-trivial and difficult to reproduce~\cite{zhou2005pfp,Csigi2017}, and enrichment appears to impose an additional constraint. In particular, although enrichment is frequently observed in networks with prominent hubs, the presence of hubs alone does not guarantee it. This is illustrated by BA networks and by CM networks with $\gamma>3$, which display heterogeneous degree distributions and high-degree nodes while showing no enrichment. Thus, enrichment is not a trivial consequence of degree heterogeneity or hub dominance. Among the models analyzed here, only the $\mathbb{S}^1$ model reproduces both RC organization and enrichment simultaneously, capturing the SRC behavior observed in some real networks.

\section*{Network effects of core enrichment}
The results above show that some networks display enrichment and/or RC organization, whereas others do not. Here we examine how these different forms of core organization are reflected in the structural relevance of high-degree nodes.

For the analyses in this section, we define a fixed high-degree core as the top $10\%$ of nodes ranked by degree, a criterion commonly used to identify the most highly connected subset of a network~\cite{vandenheuvel2011richclub}. We perform edge-concentration and core-perturbation analyses.

\paragraph{Edge concentration.} 
\begin{figure}[h!]
\centering
\includegraphics[width=0.85\linewidth]{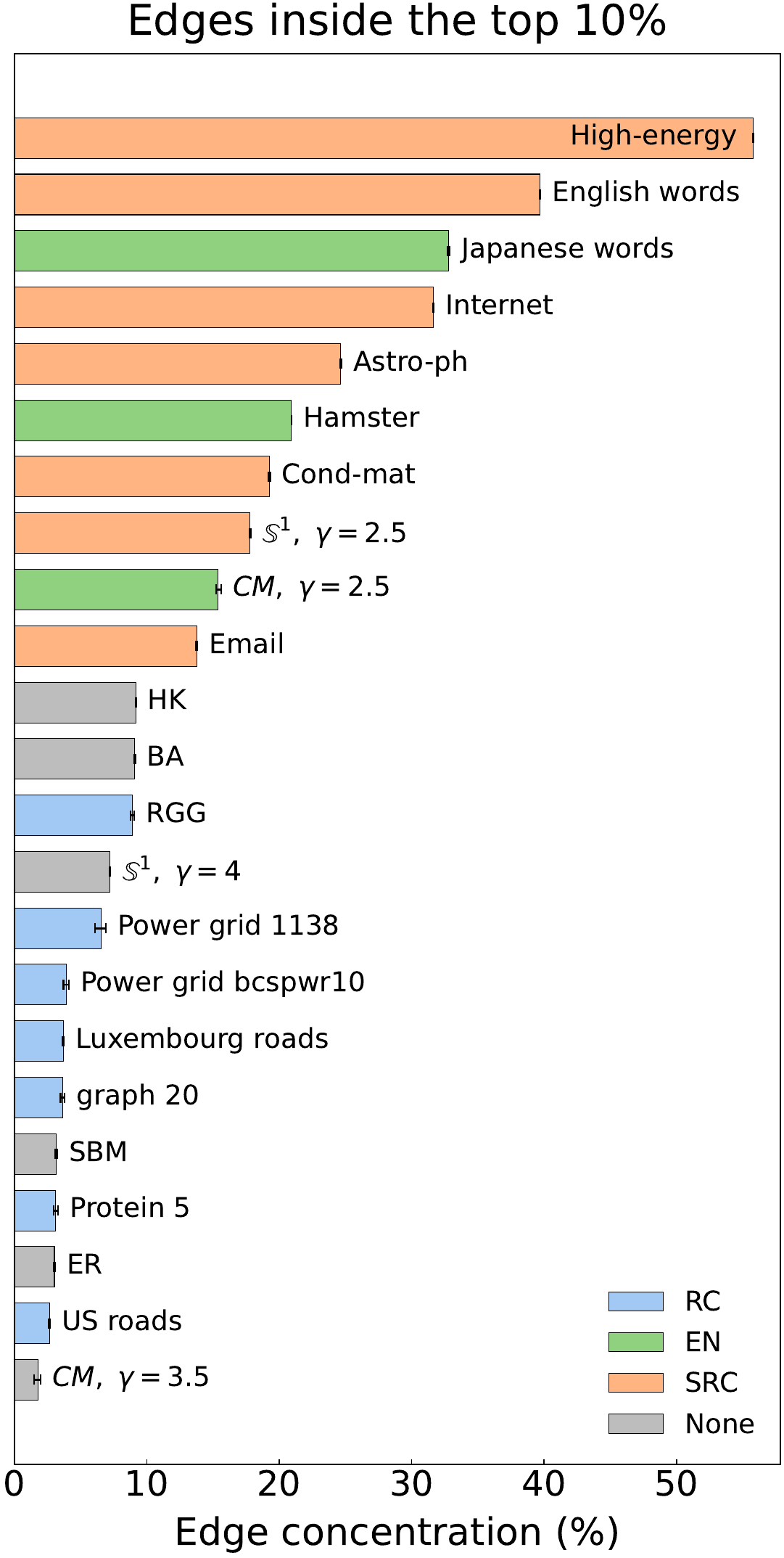}
\caption{
\textbf{Edge concentration within high-degree cores.}
Fraction of edges in the original network contained within the top $10\%$ core for empirical and synthetic networks analyzed in Figs.~\ref{fig:RC_realnet} and~\ref{fig:RC_synthetic}. Bars show the mean edge concentration, while error bars indicate the range resulting from different selections of the core when nodes are tied at the degree cutoff. Colors indicate the core-organization regime: Super Rich Club (SRC, orange), enrichment without RC organization (EN, green), RC organization without enrichment (RC, blue), and networks displaying neither property (None, gray). Enriched networks (SRC and EN) generally concentrate a larger fraction of their edges within the high-degree core than networks without enrichment.
}
\label{fig:core_metrics}
\end{figure}
We first quantify how strongly network connectivity is concentrated within the high-degree core by computing the fraction of all edges in the original network whose two endpoints belong to this subset. Notice that edge concentration is directly proportional to enrichment. For a filtered subgraph containing a fraction \(N_{>k_t}/N\) of the original nodes,
$
\frac{E_{>k_t}}{E}=\epsilon(k_t)\frac{N_{>k_t}}{N}.
$
Therefore, for cores containing the same fraction of nodes, comparing edge concentration is equivalent to comparing enrichment. We thus evaluate edge concentration at a fixed core size, providing a direct comparison across networks that is not affected by differences in the fraction of nodes retained at a given degree threshold.

The results, shown in Fig.~\ref{fig:core_metrics}, reveal a clear difference between networks with and without enrichment. Enriched networks concentrate a large fraction of their edges within this top-$10\%$ core, indicating that a small set of high-degree nodes effectively monopolizes a significant share of the network's connectivity, even for the ones  not exhibiting RC. By contrast, in RC networks without enrichment, the club accounts for only a modest fraction of the network's total links. Thus, enrichment distinguishes rich clubs that are significant in relative terms of density from those whose cores also exhibit greater internal connectivity than the original network.

\begin{figure*}[th!]
    \centering
    \includegraphics[width=0.85\linewidth]{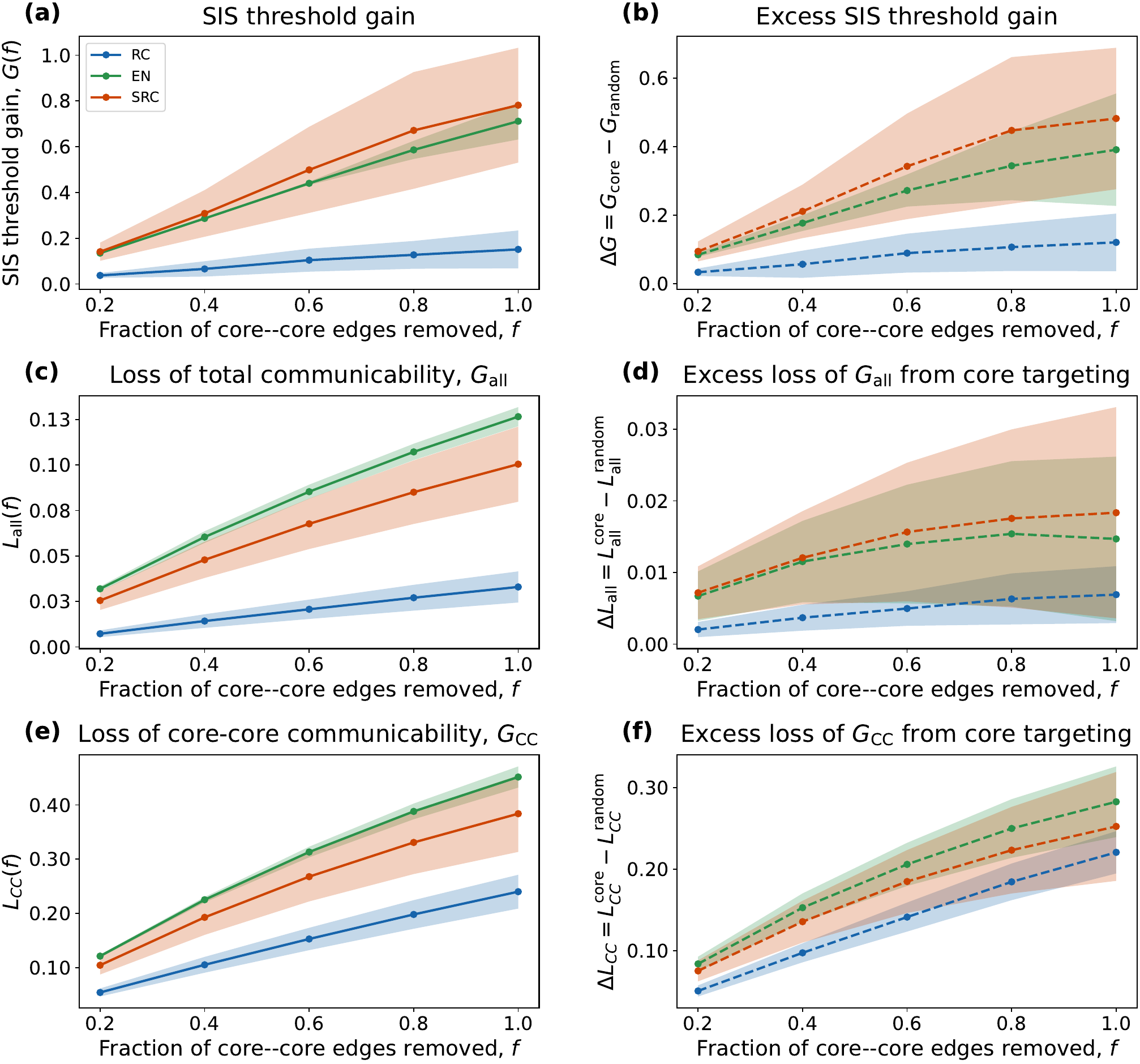}
    \caption{
    \textbf{Effect of disrupting core--core connectivity on spectral spreading and communicability of real networks.}
    A fraction $f$ of edges within the top-$10\%$ high-degree core is removed uniformly at random. 
    \textbf{(a)} Relative increase in the quenched mean-field SIS threshold, $G(f)$.
    \textbf{(b)} Excess SIS threshold gain relative to random edge removal,
    \textbf{(c)} Relative loss of total communicability, $L_{\mathrm{all}}(f)$.
    \textbf{(d)} Excess total-communicability loss relative to random edge removal.
    \textbf{(e)} Relative loss of communicability between nodes of the original high-degree core, $L_{CC}(f)$.
    \textbf{(f)} Corresponding excess loss.
  Solid lines (left panels) and dashed lines (right panels) represent the mean response across the real networks within each structural regime (RC, EN, and SRC). Shaded regions represent one standard deviation above and below the mean across networks. The analysis includes the empirical networks presented in Fig.\ref{fig:RC_realnet} that exhibit one of the three core-organization regimes.
    }
    \label{fig:core_perturbation}
\end{figure*}

\paragraph{Spectral properties and spreading capacity.}
As shown above, enriched networks concentrate a substantial fraction of their connectivity within the high-degree core. To test whether this connectivity is also dynamically relevant, we progressively disrupt the edges connecting nodes within the top-$10\%$ degree-ranked core and monitor the resulting change in the spectral threshold of the susceptible--infected--susceptible (SIS) process across the empirical networks analyzed in Fig.~\ref{fig:RC_realnet}.

Within the quenched mean-field approximation, the epidemic threshold is $\lambda_c^{\mathrm{QMF}}=1/\Lambda_{\max}(A),$ where $\Lambda_{\max}(A)$ is the leading eigenvalue of the adjacency matrix. We quantify its response to the removal of a fraction $f$ of core--core edges through
\begin{equation}
G(f)
=
\frac{\lambda_c(f)-\lambda_c(0)}
{\lambda_c(0)}.
\end{equation}

Removing core--core edges produces a markedly stronger spectral response in enriched networks (Fig.~\ref{fig:core_perturbation}a). At $f=1$, the median threshold increase reaches approximately $0.78$ in SRC networks, compared with only $0.16$ in RC networks without enrichment. Networks displaying enrichment without RC organization show a response closely comparable to SRC networks. Thus, despite their different RC properties, the two enriched regimes exhibit a similar dependence of the leading spectral mode on core connectivity.

To determine whether this response simply reflects edge removal, we compare core targeting with the removal of the same number of randomly selected edges from the whole network. Core targeting consistently produces the larger spectral effect (Fig.~\ref{fig:core_perturbation}b). The excess gain, $\Delta G=G_{\mathrm{core}}-G_{\mathrm{random}}$, is substantially larger in SRC and EN networks than in RC networks. These results indicate that the enhanced spectral sensitivity is primarily associated with enrichment, rather than with RC ordering alone.

\paragraph{Effects on communicability.}
\label{sec:communicability}
We next ask whether the influence of enriched cores extends beyond the leading spectral mode. To this end, we analyze network communicability, 
$
\mathcal{G}
=
\exp(\beta A),
$
which accounts for the multiplicity of direct and indirect walks between
nodes~\cite{estrada2008communicability,estrada2010network,
estrada2012physics} for the real networks shown in Fig. \ref{fig:RC_realnet}. We quantified communicability both across the whole network (all) and among nodes of the high-degree core (CC). See details in Methods. Using the same core-perturbation protocol described above, we measured the relative loss of communicability as
\begin{equation}
L_X(f)
=
1-
\frac{G_X(f)}{G_X(0)}
\qquad
X\in\{CC,\mathrm{all}\},
\end{equation}
and the excess loss produced by specifically targeting the high-degree core as
\begin{equation}
\Delta L_X(f)
=
L_X^{\mathrm{core}}(f)
-
L_X^{\mathrm{random}}(f).
\end{equation}

Removing core--core edges decreases communicability in all three
structural regimes, but the response is consistently stronger in enriched
networks (Fig.~\ref{fig:core_perturbation}c,e). At complete
core disruption, the median loss of total communicability is approximately
$0.13$ for EN networks and $0.10$ for SRC networks, compared with only
$0.03$ for RC networks. The corresponding losses in communicability
between core nodes are approximately $0.45$, $0.38$, and $0.24$,
respectively.

As in the spectral analysis, targeting core--core edges produces a larger
effect than removing the same number of edges randomly from the network
(Fig.~\ref{fig:core_perturbation}d,f). The excess loss in total
communicability is substantially larger for En and SRC networks than for
RC networks, whereas differences between regimes are less pronounced when
communicability is restricted to pairs of core nodes. Thus, enrichment is
associated with a stronger contribution of high-degree-core connectivity
to communication across the network, independently of RC organization.

We performed two additional structural analyses, reported in the Supplementary Information, which show that enriched networks tend to display smaller diameters and lower modularity than networks without enrichment.

\section*{Discussion and Conclusions}
RC organization identifies whether high-degree nodes are more densely connected than expected under a degree-preserving null model. Here, we introduce the term \emph{enrichment} to describe the regime in which the nested sequence of subgraphs produced under degree-threshold filtration displays an average degree exceeding that of the original network. Enrichment probes a property not captured by the normalized RC coefficient: whether progressively selecting higher-degree nodes leads to a core that is more cohesive than the original network, rather than denser than its randomized counterpart.

Across real and synthetic networks, RC organization and enrichment are neither equivalent nor mutually implied: we find networks with RC ordering but no enrichment, networks with enrichment but no RC effect, and networks---the SRC case---in which both coexist. This separation matters because a large RC coefficient does not guarantee a cohesive or dynamically dominant core: in sparse networks, the richest nodes can be more densely connected than expected while still forming a subgraph with lower mean degree than the network as a whole.
Conversely, a core can have higher average internal degree than the original network without exhibiting any significant density excess relative to its degree-preserving null model.

The structural and spreading analyses also show differences between enriched and non-enriched cores. Enriched networks (SRC and En) concentrate a substantially larger fraction of the network's edges within the highest-degree core than RC networks without enrichment, and also display smaller diameters and lower modularity. Moreover, the integrity of this core has stronger consequences both for spectral spreading capacity and for walk-based connectivity. Removing core--core edges raises the SIS spectral threshold much more strongly in enriched networks than in RC networks without enrichment, and produces larger losses of communicability, particularly at the scale of the whole network. In both analyses, targeting core--core edges has a stronger effect than removing the same number of edges randomly from the network. The similar qualitative response of En and SRC networks further suggests that these effects are associated with enrichment. By contrast, RC networks without enrichment not only concentrate fewer edges within the high-degree core, but are also less sensitive to the disruption of its internal connectivity, especially at the global level.  

These results emphasize that RC organization and enrichment capture complementary aspects of core organization. The RC coefficient identifies whether high-degree nodes are more densely connected than expected from a degree-preserving null model, whereas the enrichment coefficient reflects whether the actual connectivity per node becomes larger than in the original network as the network is filtered toward the core. According to our results, enrichment is the property most strongly associated with dynamical relevance. However, degree-preserving null models remain essential for establishing statistical significance. Importantly, it should be noted that the normalized enrichment coincides with the normalized rich club metric. Then, enrichment alone identifies a strongly interconnected core while SRC additionally establishes that this organization is stronger than expected from the network’s degree sequence alone.

The enrichment coefficient therefore complements the rich-club coefficient by distinguishing these different forms of core organization. Detecting SRC indicates a particularly strong form of core organization, relevant for identifying selective hub–hub organization beyond that expected from the degree sequence alone, for example when seeking evidence of an intentionally or evolutionarily favored backbone. 

\section*{Methods}

\subsection*{Dataset Features}
\label{sec:data}

\begin{table*}[thbp]
\centering
\setlength{\tabcolsep}{4pt}
\begin{tabular}{lccccccc}
\toprule
\textbf{Network} & \textbf{$N$} & \textbf{Mean Degree} & \textbf{Density [$10^{-3}$]} & \textbf{Degree Hubs} & \textbf{$\phi$} & \textbf{$\epsilon$} & \textbf{SRC} \\
\midrule
\rowcolor{red!5}
astro-ph ~\cite{NewmanColabNet}& 14845 & 16.120 & 1.086 & [360, 353, 329] & {\color{teal} yes} & {\color{teal} yes} & {\color{teal} yes}   \\
\rowcolor{red!5}
COND-MAT ~\cite{snap} & 23133 & 8.078 & 0.349 & [279, 252, 201] & {\color{teal} yes} & {\color{teal} yes} & {\color{teal} yes}  \\
\rowcolor{red!5}
Email ~\cite{snap} & 986 & 32.584 & 33.080 & [345, 232, 231] & {\color{teal} yes} & {\color{teal} yes} & {\color{teal} yes}   \\
\rowcolor{red!5}
ca-HepPh ~\cite{snap} & 12006 & 19.738 & 1.644 & [491, 486, 482] & {\color{teal} yes} & {\color{teal} yes} & {\color{teal} yes}   \\
\rowcolor{red!5}
internet ~\cite{internet} & 23748 & 4.919 & 0.207 & [2778, 2281, 1669] & {\color{teal} yes} & {\color{teal} yes} & {\color{teal} yes}   \\
\rowcolor{red!5}
EnglishWords~\cite{adj}  & 7377 & 11.985 & 1.625 & [2568, 1946, 1764] & {\color{teal} yes} & {\color{teal} yes} & {\color{teal} yes}   \\
\midrule
\rowcolor{cyan!5}
Protein (graph\_20) & 5748 & 4.964 & 0.864 & [11, 11, 11] & {\color{teal} yes} & {\color{red} no} & {\color{red} no}  \\
\rowcolor{cyan!5}
Protein (graph\_5)  & 4152 & 4.971 & 1.198 & [11, 10, 10] & {\color{teal} yes} & {\color{red} no} & {\color{red} no} \\
\rowcolor{cyan!5}
power\_1138bus~\cite{power-roads} & 1138 & 2.562 & 2.254 & [17, 14, 13] & {\color{teal} yes} & {\color{red} no} & {\color{red} no}  \\
\rowcolor{cyan!5}
power\_bcspwr10~\cite{power-roads}& 5300 & 3.121 & 0.589 & [13, 13, 12] & {\color{teal} yes} & {\color{red} no} & {\color{red} no}  \\
\rowcolor{cyan!5}
US roads~\cite{power-roads}  & 129164 & 2.562 & 0.020 & [7, 6, 6] & {\color{teal} yes} & {\color{red} no} & {\color{red} no}  \\
\rowcolor{cyan!5}
Luxembourg roads~\cite{power-roads}  & 114599 & 2.088 & 0.018 & [6, 5, 5] & {\color{teal} yes} & {\color{red} no} & {\color{red} no}  \\
\midrule
\rowcolor{green!4}
Hamster~\cite{power-roads} & 1858 & 13.492 & 7.265 & [272, 222, 169] & {\color{red} no} & {\color{teal} yes} & {\color{red} no}   \\
\rowcolor{green!4}
JapaneseWords~\cite{adj} & 2704 & 5.916 & 2.189 & [725, 560, 387] & {\color{red} no} & {\color{teal} yes} & {\color{red} no}   \\
\bottomrule
\end{tabular}
\caption{Summary of structural properties and classification of the analyzed empirical networks. The table reports system size $N$, mean degree, density, the three highest node degrees, and the presence of RC organization ($\phi$), enrichment ($\epsilon$) or Super Rich-Club (SRC) behavior.
}\label{tab:networks}
\end{table*}

We analyze publicly available datasets, with detailed descriptions provided in the Supplementary Information. All networks are represented as simple, unweighted, and undirected graphs, with self-loops and repeated edges removed prior to the analysis. Table~\ref{tab:networks} summarizes the main structural properties of the empirical networks, including network size, mean degree, density, highest node degrees, and their classification according to RC organization, enrichment, and SRC behavior.

\subsection*{Synthetic network models}

We analyzed synthetic networks generated with the Erd\H{o}s--R\'enyi (ER)
model~\cite{erdos1959random}, the Barab\'asi--Albert (BA) model~\cite{barabasi1999emergence}, the Holme--Kim (HK) model~\cite{holme2002growing}, a stochastic block model (SBM)~\cite{holland1983sbm}, random geometric graphs (RGG)~\cite{penrose2003random}, the configuration model (CM)~\cite{molloy1995critical}, and the $\mathbb{S}^1$ model~\cite{serrano2008selfsimilarity}.

For the $\mathbb{S}^1$ networks, mean degree $\langle k\rangle=10$, and inverse temperature $\beta=2$, considering degree-distribution exponents $\gamma=2.5$ and $\gamma=4$. The corresponding CM networks were obtained by degree-preserving randomization of these $\mathbb{S}^1$ networks.

The remaining synthetic networks were generated using the NetworkX Python library~\cite{hagberg2008networkx}. ER and BA networks had mean degree $\langle k\rangle\simeq6$. HK networks were generated with $m=3$ edges
added per new node, corresponding to $\langle k\rangle\simeq6$, and
triad-formation probability $p=0.5$. SBM networks consisted of four equally sized groups, with intra-group and inter-group connection probabilities $p_{\rm in}$ and $p_{\rm out}$, respectively, chosen such that $p_{\rm in}/p_{\rm out}=3$, and with $\langle k\rangle\simeq6$. RGG networks were generated in two spatial dimensions with $\langle k\rangle\simeq6$.

Further details on the generation of the synthetic networks are provided
in the Supplementary Information.

\subsection*{Structural condition for enrichment}
\label{sec:enrichment_condition}
We derive here the condition under which the average degree of the network increases after removing all nodes with degree $k < k_{\text{threshold}}$.

Let $E$ and $N$ be the number of edges and nodes in the original network, respectively. 
The average degree is therefore
\begin{equation}
\langle k \rangle = \frac{2E}{N}.
\end{equation}
Let $r$ be the number of nodes removed, and let $R$ denote the set of removed nodes. 
We denote by $\sum_{r \in R} d$ the sum of the degrees of the removed nodes, and by $m_r$ the number of edges among those nodes.

When removing nodes, all edges incident to them are deleted. However, edges among removed nodes are counted twice in the sum of degrees and must therefore be corrected. 
The number of edges removed is thus
\begin{equation}
\sum_{r \in R} d - m_r.
\end{equation}
The remaining number of edges is
\begin{equation}
E_s = E - \left(\sum_{r \in R} d - m_r \right),
\end{equation}
and the remaining number of nodes is $N - r$. 
Therefore, the average degree after removal is
\begin{equation}
\begin{aligned}
\langle k \rangle_s
&=
\frac{2}{N-r}
\left[
E-
\left(
\sum_{r\in R}d-m_r
\right)
\right].
\end{aligned}
\end{equation}

The condition for enrichment is $\langle k \rangle_s > \langle k \rangle$, which gives
\begin{equation}
\begin{aligned}
\frac{2}{N-r}
\left[
E-
\left(
\sum_{r\in R}d-m_r
\right)
\right]
>
\frac{2E}{N}.
\end{aligned}
\end{equation}
Multiplying both sides by $N(N - r)$:
\begin{equation}
\begin{aligned}
2N
\left[
E-
\left(
\sum_{r\in R}d-m_r
\right)
\right]
>
2E(N-r).
\end{aligned}
\end{equation}
Expanding both sides:
\begin{equation}
\begin{aligned}
2EN
-
2N
\left(
\sum_{r\in R}d-m_r
\right)
>
2EN-2Er.
\end{aligned}
\end{equation}
Canceling identical terms:
\begin{equation}
\begin{aligned}
2N
\left(
\sum_{r\in R}d-m_r
\right)
<
2Er.
\end{aligned}
\end{equation}
Dividing both sides by $rN$:
\begin{equation}
2\frac{\sum_{r \in R} d - m_r}{r}
<
\frac{2E}{N}.
\end{equation}
We now rewrite the left-hand side in terms of averages. 
The average degree of the removed nodes is
\begin{equation}
\langle k \rangle_r
=
\frac{1}{r}
\sum_{r \in R} d,
\end{equation}
so that
\begin{equation}
\begin{aligned}
\frac{\sum_{r \in R} d - m_r}{r}
&=
\langle k \rangle_r
-
\frac{m_r}{r}.
\end{aligned}
\end{equation}
Substituting, we obtain
\begin{equation}
2\left(
\langle k \rangle_r
-
\frac{m_r}{r}
\right)
<
\langle k \rangle.
\end{equation}

This is the final condition for enrichment.

\medskip

An equivalent interpretation can be obtained by introducing the internal average degree of the removed nodes,
\begin{equation}
\langle k \rangle_{\mathrm{int}}
=
\frac{2m_r}{r}.
\end{equation}
The condition becomes
\begin{equation}
\begin{aligned}
\langle k \rangle_r
-
\frac{\langle k \rangle_{\mathrm{int}}}{2}
<
\frac{\langle k \rangle}{2}.
\end{aligned}
\end{equation}

Thus, enrichment occurs when the corrected average degree of the removed nodes is smaller than half the average degree of the original network.

\subsection*{Perturbation protocol}
\label{sec:perturbation}
For each network, we defined the high-degree core as the set $C_0$
containing the top $10\%$ of nodes ranked by degree in the unperturbed
network. The set of core nodes was kept fixed throughout the perturbation
procedure.

Starting from the original adjacency matrix $A_0$, we progressively
disrupted the connectivity within the core by removing uniformly at random
a fraction $f$ of the edges whose two endpoints belong to $C_0$. The
resulting adjacency matrix is denoted by $A_f$. All edges not connecting
two core nodes were left unchanged.

For each value of $f$, the perturbation was repeated over $100$
independent realizations. As a control, we performed an equivalent random
perturbation in which the same number of edges removed in the
core-targeted experiment was instead selected uniformly at random from
the entire network. This ensures that differences between targeted and
random perturbations arise from the location of the removed edges rather
than from the number of edges removed.

\paragraph{Spectral SIS threshold.}
We quantified the effect of core disruption on spreading capacity using
the quenched mean-field approximation of the susceptible--infected--susceptible (SIS) model. For a network with adjacency matrix $A_f$, the
spectral epidemic threshold is
\begin{equation}
\lambda_c(f)
=
\frac{1}{\Lambda_{\max}(A_f)},
\end{equation}
where $\Lambda_{\max}(A_f)$ is the largest eigenvalue of the adjacency
matrix~\cite{castellano2010thresholds,chakrabarti2008epidemic}.

The relative change in the spectral threshold with respect to the
unperturbed network was quantified as
\begin{equation}
G(f)
=
\frac{\lambda_c(f)-\lambda_c(0)}
{\lambda_c(0)}.
\end{equation}

To isolate the effect associated with targeting the
high-degree core, we defined the excess spectral response
\begin{equation}
\Delta G(f)
=
G_{\mathrm{core}}(f)
-
G_{\mathrm{random}}(f),
\end{equation}
where $G_{\mathrm{core}}(f)$ corresponds to the removal of core--core
edges and $G_{\mathrm{random}}(f)$ to the removal of the same number of
edges selected randomly from the whole network.

\paragraph{Network communicability.}
For an undirected network with adjacency matrix $A$, communicability
between nodes $i$ and $j$ is defined through the matrix exponential
\begin{equation}
\mathcal{G}_{ij}(\beta)
=
\left[\exp(\beta A)\right]_{ij}
=
\sum_{\ell=0}^{\infty}
\frac{\beta^\ell(A^\ell)_{ij}}{\ell!},
\end{equation}
where $(A^\ell)_{ij}$ gives the number of walks of length $\ell$
connecting nodes $i$ and $j$~\cite{estrada2008communicability,
estrada2010network,estrada2012physics}.

For each network, the communicability scale was fixed using the
unperturbed adjacency matrix $A_0$ as
\begin{equation}
\beta_0
=
\frac{1}{\Lambda_{\max}(A_0)}.
\end{equation}
The same value of $\beta_0$ was used for the original and all perturbed
networks,
\begin{equation}
\mathcal{G}(f)
=
\exp(\beta_0 A_f),
\end{equation}
so that changes in communicability along the perturbation sequence arise
solely from changes in network structure rather than from a rescaling of
walk lengths.

We evaluated communicability at two structural scales. The normalized
communicability associated with pairs of nodes belonging to the original
core was defined as
\begin{equation}
G_{CC}(f)
=
\frac{
\mathbf{c}^{T}
\exp(\beta_0 A_f)
\mathbf{c}
}
{N_C^2},
\end{equation}
where $\mathbf{c}$ is the indicator vector of $C_0$ and
$N_C=|C_0|$. Walks contributing to this quantity are not restricted to
the core and may traverse any part of the network.

The normalized total communicability of the full network was defined as
\begin{equation}
G_{\mathrm{all}}(f)
=
\frac{
\mathbf{1}^{T}
\exp(\beta_0 A_f)
\mathbf{1}
}
{N^2},
\end{equation}
where $\mathbf{1}$ is the all-ones vector.

\section*{Acknowledgements}
We thank Mari\'an Bogu\~n\'a for helpful comments. L.B. acknowledges support from an FPU21/03183 grant from the Spanish Government. M.\'A.S acknowledge support from Grant No. PID2022-137505NB-C22 funded by MCIN/AEI/10.13039/501100011033 and by ERDF/EU and Grant No. PID2025-169261NB-C21 funded by MICIU/AEI/10.13039/501100011033.


\begin{thebibliography}{34}

\bibitem{albert2002statistical}
Albert, R. \& Barab\'asi, A.-L. Statistical mechanics of complex networks. \textit{Rev. Mod. Phys.} \textbf{74}, 47--97 (2002).

\bibitem{song2006origins}
Song, C., Havlin, S. \& Makse, H. A. Origins of fractality in the growth of complex networks. \textit{Nat. Phys.} \textbf{2}, 275--281 (2006).

\bibitem{castellano2012competing}
Castellano, C. \& Pastor-Satorras, R. Competing activation mechanisms in epidemics on networks. \textit{Sci. Rep.} \textbf{2}, 371 (2012).

\bibitem{castellano2017relating}
Castellano, C. \& Pastor-Satorras, R. Relating topological determinants of complex networks to their spectral properties: structural and dynamical effects. \textit{Phys. Rev. X} \textbf{7}, 041024 (2017).

\bibitem{zhou2004rich}
Zhou, S. \& Mondrag\'on, R. J. The rich-club phenomenon in the Internet topology. \textit{IEEE Commun. Lett.} \textbf{8}, 180--182 (2004).

\bibitem{colizza2006detecting}
Colizza, V., Flammini, A., Serrano, M. A. \& Vespignani, A. Detecting rich-club ordering in complex networks. \textit{Nat. Phys.} \textbf{2}, 110--115 (2006).

\bibitem{serrano2008rich}
Serrano, M. A. Rich-club vs rich-multipolarization phenomena in weighted networks. \textit{Phys. Rev. E} \textbf{78}, 026101 (2008).

\bibitem{opsahl2008prominence}
Opsahl, T., Colizza, V., Panzarasa, P. \& Ramasco, J. J. Prominence and control: the weighted rich-club effect. \textit{Phys. Rev. Lett.} \textbf{101}, 168702 (2008).

\bibitem{zlatic2009rich}
Zlatic, V. \textit{et al.} On the rich-club effect in dense and weighted networks. \textit{Eur. Phys. J. B} \textbf{67}, 271--275 (2009).

\bibitem{mcauley2007richclub}
McAuley, J. J., da Fontoura Costa, L. \& Caetano, T. S. The rich-club phenomenon across complex network hierarchies. \textit{Appl. Phys. Lett.} \textbf{91}, 084103 (2007).

\bibitem{pedreschi2022temporal}
Pedreschi, N., Battaglia, D. \& Barrat, A. The temporal rich club phenomenon. \textit{Nat. Phys.} \textbf{18}, 931--938 (2022).

\bibitem{ma2015anatomy}
Ma, A., Mondrag\'on, R. J. \& Latora, V. Anatomy of funded research in science. \textit{Proc. Natl Acad. Sci. USA} \textbf{112}, 14760--14765 (2015).

\bibitem{nakajima2023higher}
Nakajima, K., Shudo, K. \& Masuda, N. Higher-order rich-club phenomenon in collaborative research grant networks. \textit{Scientometrics} \textbf{128}, 2429--2446 (2023).

\bibitem{vandenheuvel2011richclub}
van den Heuvel, M. P. \& Sporns, O. Rich-club organization of the human connectome. \textit{J. Neurosci.} \textbf{31}, 15775--15786 (2011).

\bibitem{nigam2016rich}
Nigam, S. \textit{et al.} Rich-club organization in effective connectivity among cortical neurons. \textit{J. Neurosci.} \textbf{36}, 670--684 (2016).

\bibitem{zheng2024rich}
Zheng, J. \textit{et al.} Rich-club organization of whole-brain spatio-temporal multilayer functional connectivity networks. \textit{Front. Neurosci.} \textbf{18}, 1405734 (2024).

\bibitem{riedel2022}
Riedel, L., van den Heuvel, M. P. \& Markett, S. Trajectory of rich club properties in structural brain networks. \textit{Hum. Brain Mapp.} \textbf{43}, 4239--4253 (2022).

\bibitem{vandenheuvel2013network}
van den Heuvel, M. P. \& Sporns, O. Network hubs in the human brain. \textit{Trends Cogn. Sci.} \textbf{17}, 683--696 (2013).

\bibitem{erdos1959random}
Erd\H{o}s, P. \& R\'enyi, A. On random graphs. \textit{Publ. Math.} \textbf{6}, 290--297 (1959).

\bibitem{molloy1995critical}
Molloy, M. \& Reed, B. A critical point for random graphs with a given degree sequence. \textit{Random Struct. Algorithms} \textbf{6}, 161--180 (1995).

\bibitem{serrano2008selfsimilarity}
Serrano, M. A., Krioukov, D. \& Bogu\~n\'a, M. Self-similarity of complex networks and hidden metric spaces. \textit{Phys. Rev. Lett.} \textbf{100}, 078701 (2008).

\bibitem{barabasi1999emergence}
Barab\'asi, A.-L. \& Albert, R. Emergence of scaling in random networks. \textit{Science} \textbf{286}, 509--512 (1999).

\bibitem{holme2002growing}
Holme, P. \& Kim, B. J. Growing scale-free networks with tunable clustering. \textit{Phys. Rev. E} \textbf{65}, 026107 (2002).

\bibitem{holland1983sbm}
Holland, P. W., Laskey, K. B. \& Leinhardt, S. Stochastic blockmodels: first steps. \textit{Soc. Netw.} \textbf{5}, 109--137 (1983).

\bibitem{penrose2003random}
Penrose, M. \textit{Random Geometric Graphs} (Oxford Univ. Press, 2003).

\bibitem{zhou2005pfp}
Zhou, S. \& Mondrag\'on, R. J. Accurately modeling the internet topology. \textit{Phys. Rev. E} \textbf{70}, 066108 (2004).

\bibitem{Csigi2017}
Csigi, M. \textit{et al.} Geometric explanation of the rich-club phenomenon in complex networks. \textit{Sci. Rep.} \textbf{7}, 1730 (2017).

\bibitem{estrada2008communicability}
Estrada, E. \& Hatano, N. Communicability in complex networks. \textit{Phys. Rev. E} \textbf{77}, 036111 (2008).

\bibitem{estrada2010network}
Estrada, E. \& Higham, D. J. Network properties revealed through matrix functions. \textit{SIAM Rev.} \textbf{52}, 696--714 (2010).

\bibitem{estrada2012physics}
Estrada, E., Hatano, N. \& Benzi, M. The physics of communicability in complex networks. \textit{Phys. Rep.} \textbf{514}, 89--119 (2012).

\bibitem{NewmanColabNet}
Newman, M. E. J. Scientific collaboration networks. I. Network construction and fundamental results. \textit{Phys. Rev. E} \textbf{64}, 016131 (2001).

\bibitem{snap}
Leskovec, J., Kleinberg, J. \& Faloutsos, C. Graph evolution: densification and shrinking diameters. \textit{ACM Trans. Knowl. Discov. Data} \textbf{1}, 2 (2007).

\bibitem{internet}
Bogu\~n\'a, M., Papadopoulos, F. \& Krioukov, D. Sustaining the Internet with hyperbolic mapping. \textit{Nat. Commun.} \textbf{1}, 62 (2010).

\bibitem{adj}
Milo, R. \textit{et al.} Superfamilies of evolved and designed networks. \textit{Science} \textbf{303}, 1538--1542 (2004).

\bibitem{power-roads}
Rossi, R. A. \& Ahmed, N. K. The Network Data Repository with Interactive Graph Analytics and Visualization. In \textit{Proc. Twenty-Ninth AAAI Conf. Artificial Intelligence} 4292--4293 (2015).

\bibitem{castellano2010thresholds}
Castellano, C. \& Pastor-Satorras, R. Thresholds for epidemic spreading in networks. \textit{Phys. Rev. Lett.} \textbf{105}, 218701 (2010).

\bibitem{chakrabarti2008epidemic}
Chakrabarti, D., Wang, Y., Wang, C., Leskovec, J. \& Faloutsos, C. Epidemic thresholds in real networks. \textit{ACM Trans. Inf. Syst. Secur.} \textbf{10}, 1 (2008).

\bibitem{hagberg2008networkx}
Hagberg, A. A., Schult, D. A. \& Swart, P. J. Exploring network structure, dynamics, and function using NetworkX. In \textit{Proc. 7th Python in Science Conf.} 11--15 (2008).

\end{thebibliography}
\end{document}